\documentclass[twocolumn]{jpsj2} 
\title{$^{59}$Co and $^{75}$As NMR Investigation of Electron-Doped High $T_c$ Superconductor BaFe$_{1.8}$Co$_{0.2}$As$_{2}$ ($T_{c} = 22$~K)}

\author{Fanlong \textsc{Ning}$^{1}$, Kanagasingham \textsc{Ahilan}$^{1}$, Takashi  \textsc{Imai}$^{1,2}$\thanks{E-mail: imai@mcmaster.ca}, Athena S.  \textsc{Sefat}$^{3}$, Ronying  \textsc{Jin}$^{3}$, Michael A.  \textsc{McGuire}$^{3}$, Brian C.  \textsc{Sales}$^{3}$, and David  \textsc{Mandrus}$^{3}$}

\inst{$^{1}$Department of Physics and Astronomy, McMaster University, Hamilton, Ontario L8S4M1, Canada\\
$^{2}$Canadian Institute for Advanced Research, Toronto, Ontario M5G1Z8, Canada \\
$^{3}$Materials Science and Technology Division, Oak Ridge National Laboratory, TN 37831, USA}

\abst{We report an NMR investigation of the superconductivity in BaFe$_{2}$As$_{2}$ induced by Co doping ($T_{c} = 22$~K).  We demonstrate that Co atoms form an alloy with Fe atoms and donate carriers without creating localized moments.  Our finding strongly suggests that the underlying physics of iron-pnictide superconductors is quite different from the widely accepted physical picture of high $T_c$ cuprates as doped Mott insulators.  We also show a crossover of electronic properties into a low temperature pseudo-gap phase with a pseudo-gap $\Delta_{PG}/k_{B} \sim 560$~K, where $\chi_{spin} \sim constant$ and resisitivty $\rho \propto T$. The NMR Knight shift below $T_c$ decreases for both along the c-axis and ab-plane, and is consistent with the singlet pairing scenario.}

\kword{iron pnictide superconductor, high temperature superconductivity, NMR}

\begin{document}
\maketitle

The recent discovery of  iron-pnictide high-temperature superconductors with transition temperatures as high as  $T_{c} = 26 \sim55$~K \cite{kamihara,sefat1,sm,ren,rotter} raises many fundamental questions.  One of the central issues is whether or not the underlying physics of the new iron-pnictides is similar to that of high-$T_c$ cuprate superconductors.  Certainly there are some remarkable similarities: the undoped parent phases, LaFeAsO and BaFe$_2$As$_2$, show magnetic long range order\cite{delacruz,bafeas}, and LaFeAsO$_{1-\delta}$F$_{\delta}$ \cite{kamihara} and Ba$_{1-2\delta}$K$_{2\delta}$Fe$_2$As$_2$\cite{rotter} become superconducting when the LaO and Ba charge reservoir layers donate superconducting carriers to the FeAs layers.  This situation is reminiscent of the case of the high-$T_c$ cuprates, where carrier doping from charge reservoir layers transforms the undoped Mott-insulating CuO$_2$ layers into superconducting (CuO$_2$)$^{-2+\delta}$, where $\delta$ is the carrier concentration.  

In the high-$T_c$ cuprates, one can also substitute Zn$^{2+}$ (3d$^{10}$, spin $S=0$) and Ni$^{2+}$ (3d$^{8}$, $S=1$) ions into Cu$^{2+}$ (3d$^{9}$, $S=\frac{1}{2}$) sites.  However, Zn$^{2+}$ and Ni$^{2+}$ ions induce localized magnetic moments in their vicinity, and destroy the superconductivity \cite{ZnTc}.  NMR is an ideal local probe to investigate spatially modulating spin densities induced by dopants, and earlier NMR studies in the cuprates have observed a Curie-like behavior of the induced local moments through Knight shift measurements  \cite{ishida,bobroff}.  In view of this precedent in the cuprates, the recent discoveries of superconductivity in Co doped La(Fe$_{1-\delta}$Co$_{\delta}$)AsO \cite{sefat2} and Ba(Fe$_{2-2\delta}$Co$_{2\delta}$)As$_2$ \cite{sefat3, german} came as a major surprise.  Notice that Co is located next to Fe in the periodic table, and has one extra electron.  Within the context of Mott physics, Co substitution into Fe sites should be detrimental to superconductivity in analogy with Zn substitution into Cu sites in high-$T_c$ cuprates.  Instead, {\it Co atoms appear to donate the extra electrons to Fe layers as itinerant carriers, and induce superconductivity}.  The goal of the present study is to investigate the local electronic properties at the dopant Co sites and in their vicinity, using $^{59}$Co and $^{75}$As NMR.  We show that {\it the temperature dependence of the local spin susceptibility $\chi_{Co}$ at Co sites, as measured by  $^{59}$Co NMR Knight shift $^{59}$K, is identical to that of the host Fe$_2$As$_2$ layers as determined by  $^{75}$As Knight shift $^{75}$K}.  Moreover, we find no evidence for the presence of localized moments with Curie behavior.  That is, {\it Co and Fe can form two dimensional square-lattice sheets of a superconducting alloy}. We also demonstrate that low energy spin excitations decrease with temeprature, and level off below $T^{*}\sim 100$~K.  Combined with our earlier $^{19}$F NMR discovery of analogous behavior in LaFeAsO$_{0.89}$F$_{0.11}$ \cite{ahilan1}, this implies that {\it pseudo-gap} behavior may be universal in iron-pnictide supercondcutors.  The use of a single crystal enabled us,  for the first time, to measure $^{75}K$ below $T_c$ for {\it both} along the crystal c-axis {\it and} ab-plane.  Our results are consistent with the singlet pairing scenario.  

Our single crystal sample is from the same batch used for specific heat measurements which established the bulk nature of superconductivity in BaFe$_{1.8}$Co$_{0.2}$As$_{2}$.\cite{sefat3}  We cut out a small crystal with total weight of 31~mg for our NMR measurements.  We set aside a small portion ($\sim3$~mg) of the NMR crystal for charge transport measurements.\cite{ahilan2}  The resistive superconducting transition is $T_{c} = 22$~K,\cite{sefat3,ahilan2} and an application of $B=7.7$~Tesla magnetic field reduces the midpoint of the superconducting transition to $T_c =  16.0$~K for $B \parallel c$ and $T_{c} = 18.8$~K for $B \parallel ab$.   The total weight of Co atoms in the NMR sample is only $\sim$1~mg (i.e. $\sim10^{19}$ Co atoms), but $^{59}$Co NMR signals were readily observable.  Except for the $^{75}K_{ab}$ data with $B \parallel ab$ in Fig.2, we conducted all NMR measurements with $B \parallel c$ axis.       

In Fig.1, we present field swept NMR spectra at 25~K of $^{59}$Co (nuclear spin $I = \frac{7}{2}$, nuclear gyromagnetic ratio $^{59}\gamma_{n}/2\pi = 10.054$~MHz/Tesla) and $^{75}$As ($I = \frac{3}{2}$, $^{75}\gamma_{n}/2\pi = 7.2919$~MHz/Tesla).  Each Co site gives rise to seven $I_{z}=(2m-1)/2$ to $(2m+1)/2$ transitions (where integer $m = -3$ to $+3$), separated by the c-axis component of the nuclear quadrupole interaction tensor, $^{59}\nu_{Q}^{c}$.  The $^{59}\nu_{Q}^{c}$ is proportional to the second derivative of the Coulomb potential, and sensitive to the local lattice environment.  If we assume that Co atoms substitute into Fe sites randomly in the Fe$_{1.8}$Co$_{0.2}$ layers with equal probability $p = 0.2 / 2 = 0.1$, each Co site may have $N =0$ to $4$ nearest neighbor Co sites with the probability of $P(N)=_{4}C_{N} p^{N} (1-p)^{4-N} = 0.66$~(for $N=0$), $0.29$~($N=1$), $0.05$~($N=2$), $0.004$~($N=3$), and $0.0001$~($N=4$).  We attribute the seven main peaks marked by red arrows to the $N=0$ sites with $^{59}\nu_{Q}^{c}/ ^{59}\gamma_{n} \sim 0.026$~Tesla. Two additional peaks marked by green dashed arrows should be attributed to the $m=\pm3$ transitions of the $N=1$ sites with $^{59}\nu_{Q}^{c}/ ^{59}\gamma_{n} \sim 0.034$~Tesla.  We also assign a small peak at $B\sim 7.3$~Tesla tentatively to the $m=-1$ transition of the $N=2$ sites.  The $^{75}$As NMR lineshape in Fig.3b reveals only one set of three peaks arising from $m=-1, 0, +1$ transitions, but the $m = \pm 1$ satellite transitions are very broad.  This means that the alloying of Fe and Co induces a substantial distribution in $^{75}\nu_{Q}^{c}/^{75}\gamma_{n} \sim 0.33$~Tesla.  We also observed a small unidentified peak at $B\sim 7.78$~Tesla which accounts for $\sim$6~\% of the total integrated intensity.  However, we confirmed that the whole $^{59}$Co and $^{75}$As NMR lineshapes show no noticeable changes between 25~K and 50~K, hence the small unidentified peaks in Fig.3a and 3b do not exhibit Curie-like behavior even if they are indeed $^{59}$Co and $^{75}$As NMR signals from BaFe$_{1.8}$Co$_{0.2}$As$_{2}$.  

\begin{figure}[b]
\begin{center}
\includegraphics[width=7cm]{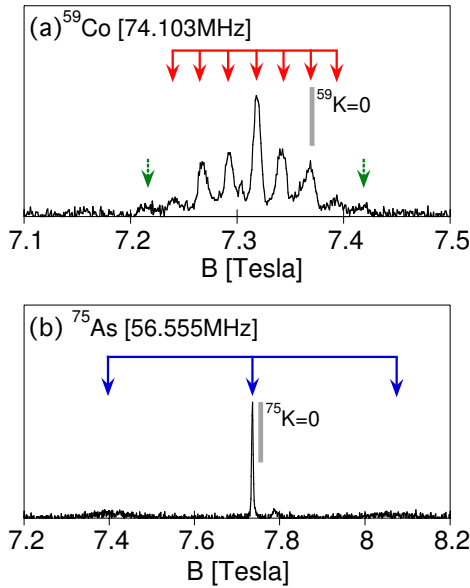}
\end{center}
\caption{(Color online) (a) Field swept $^{59}$Co NMR lineshape observed at 25~K ($B \parallel c$).  Red arrows and green dashed arrows ientify the $N=0$ and $N=1$ sites, respectively (see main text).  (b) Field swept $^{75}$As NMR line at 25~K ($B \parallel c$).   Blue arrows identify the three allowed transitions.  Vertical thick grey lines mark the location of the $m=0$ central transition if $^{59,75}K=0$.
}
\label{f1}
\end{figure}    

In Fig.2, we summarize the temperature dependence of the $^{59}$Co and $^{75}$As NMR Knight shift, $^{59}K$ and $^{75}K$.  The observed temperature dependence is qualitatively similar to that observed in undoped BaFe$_{2}$As$_{2}$ above $T_N$,\cite{fukazawa,baek} and also in superconducting LaFeAsO$_{1-\delta}$F$_{\delta}$ ($\delta \sim 0.1$).\cite{ahilan1,grafe}  The magnitude of $^{75}K$ is somewhat smaller than the paramagnetic state of BaFe$_{2}$As$_{2}$.\cite{fukazawa,baek}  We can relate $^{59}K$ and $^{75}K$ with the local spin susceptibility as follows:
\begin{subequations}
\begin{align}
^{59}K & = A \chi_{Co} +  \sum_{j=1 \sim 4} B\chi_{j} + ^{59}K_{chem},\\
^{75}K & =  \sum_{j=1 \sim 4} C \chi_{j} + ^{75}K_{chem},\
\end{align}
\end{subequations}
where $A$ represents the hyperfine coupling between the $^{59}$Co nuclear spin and the Co electron spin at the same site.  $B$ is the transferred hyperfine interaction between the $^{59}$Co nuclear spin and the electrons at the four nearest neighbor Fe or Co sites, and $\chi_j$ ($=\chi_{Co}$, or $\chi_{Fe}$ at Fe sites) represents the electron spin susceptibility at the nearest neighbor sites.  $^{59,75}K_{chem}$ is the temperature independent chemical shift.   $C$ is the transferred hyperfine coupling between the $^{75}$As nuclear spin and electrons at the four nearest neighbor Fe or Co sites.  Generally, $A<0$ for the on-site term of transition metals, and the transferred terms are positive, $B>0$ and $C>0$.  

\begin{figure}[b]
\begin{center}
\includegraphics[width=9cm]{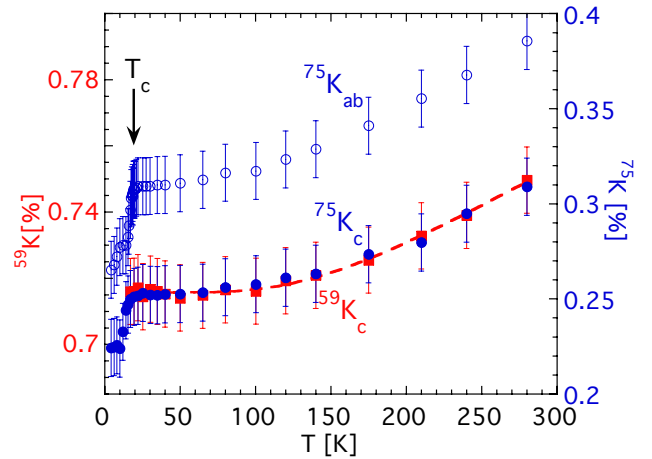}
\end{center}
\caption{(Color online) Temperature dependence of the $^{59}$Co and $^{75}$As NMR Knight shifts, $^{59}K_{c}$ (red squares), $^{75}K_{c}$ (blue circles), and $^{75}K_{ab}$ (blue open circles) in an external magnetic field $B=7.7$~Tesla applied along the c-axis or ab-plane.  The dashed curve is a fit to an activation form above $T_c$, $^{59}K = A + B \times exp(- \Delta_{PG} / k_{B}T)$ with $A=0.715\%$, $B=0.244\%$, and the pseudo gap $\Delta_{PG}/k_{B} = 560\pm150$~K.}
\label{f2}
\end{figure}

Notice that we observe only one $m=0$ central peak of $^{59}$Co and $^{75}$As lines, and hence $^{59}K$ and $^{75}K$ are single valued, despite the random substitution of Co into Fe sites.  Furthermore, the temperature dependence of $^{59}K$ and $^{75}K$ is identical, and shows no hint of Curie behavior.  These findings are in remarkable contrast to the case of the high-$T_c$ cuprates.  When one substitutes Cu$^{2+}$ ions with Li$^{+}$, Zn$^{2+}$, or Al$^{3+}$ cations in CuO$_2$ planes of the high-$T_c$ cuprates, the NMR Knight shifts observed at cation sites, as well as at nearby $^{17}$O and $^{89}$Y sites, exhibit a Curie behavior, $K \sim 1/T$.\cite{ishida,bobroff}  Such a Curie behavior arises because substituted ions induce localized magnetic moments in the essentially localized electrons at Cu sites in doped Mott insulators.  In contrast, our results imply that {\it (i) Fe and Co sites carry comparable electron spin susceptibility (i.e. $\chi_{Co} \sim \chi_{Fe}$), (ii) Co substitution into Fe sites does not induce localized magnetic moments in the vicinity of Co sites, and hence (iii) the spatial variation of local spin susceptibility $\chi_{spin}$ is small.}  This inevitably leads us to suggest that {\it the underlying physics of the iron-pnictide superconductors is quite different from the Mott physics which is generally believe to describe the high-$T_c$ cuprates.}

In Fig.3, we present the temperature dependence of the nuclear spin-lattice relaxation rates $^{59,75}(\frac{1}{T_1})$, and 
\begin{equation}
^{59,75}(\frac{1}{T_{1}T}) \sim  \sum_{\bf q} (^{59,75}\gamma_{n})^{2} |^{59,75}A({\bf q})|^{2} \frac{ \chi " ({\bf q},f)}{f},
\end{equation}
where $|^{59,75}A({\bf q})|^{2}$ is the wave-vector ${\bf q}$-dependent hyperfine form factor, $\chi " ({\bf q},f)$ is the imaginary part of the dynamical electron spin susceptibility (i.e. spin fluctuations), and $f$ is the NMR frequency ($\lesssim 10^{2}$~MHz).  Attributing the local spin density primarily to 3d orbitals at the Fe and Co sites, we may represent 
\begin{subequations}
\begin{align}
|^{59}A({\bf q})|^{2} & = |A + 2B (cos(q_{x}a) + cos(q_{y}a)) |^{2},\\
|^{75}A({\bf q})|^{2} & = |4C cos(q_{x}a/2)cos(q_{y}a/2)|^{2},\
\end{align}
\end{subequations}
where $a$ is the distance between Fe sites.  We carried out all $^{59,75}(\frac{1}{T_1})$ measurements for the central transition in an external magnetic field of $\sim 7.7$~Tesla, and the recovery of the nuclear magnetization, $M(t)$, after an inversion pulse was fit with the appropriate solutions to the Master equation.  As shown in Fig.4, the fit of $M(t)$ was good except below $\sim$10~K, where vortices induced by external magnetic field are expected to cause a distribution of $^{59,75}(\frac{1}{T_1})$.  For $^{59}$Co, we also confirmed that $^{59}(\frac{1}{T_1})$ is comparable at the satellite transitions for different types of Co sites with $N=0$, $1$ and $2$.    
\begin{figure}[b]
\begin{center}
\includegraphics[width=9.5cm]{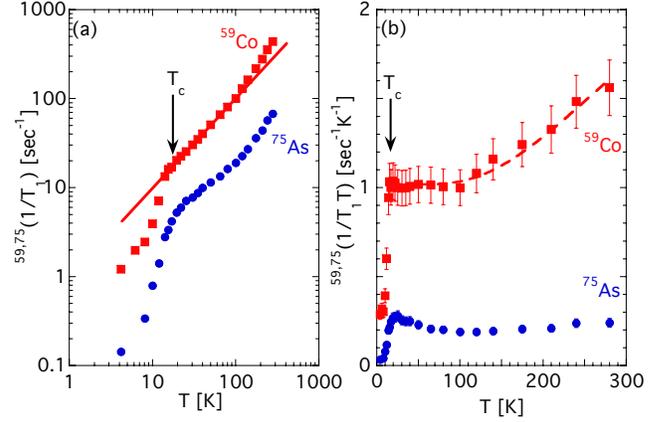}
\end{center}
\caption{(Color online)  (a) Temperature dependence of the $^{59}$Co and $^{75}$As nuclear spin lattice relaxation rate $^{59,75}(\frac{1}{T_{1}})$.  Solid line represents a $T$-linear behavior.    (b) $^{59}(\frac{1}{T_{1}T})$ and $^{75}(\frac{1}{T_{1}T})$.  The dashed curve is a fit to an activation form, $^{59}(\frac{1}{T_{1}T}) = A + B \times exp(- \Delta_{PG} / k_{B}T)$ with $A=1.0$~sec$^{-1}$K$^{-1}$, $B=4.3$~sec$^{-1}$K$^{-1}$, and the same magnitude of pseudo-gap $\Delta_{PG}/k_{B} = 560 \pm 150 $~K as in Fig.2.}
\label{f3}
\end{figure}
\begin{figure}[b]
\begin{center}
\includegraphics[width=7.5cm]{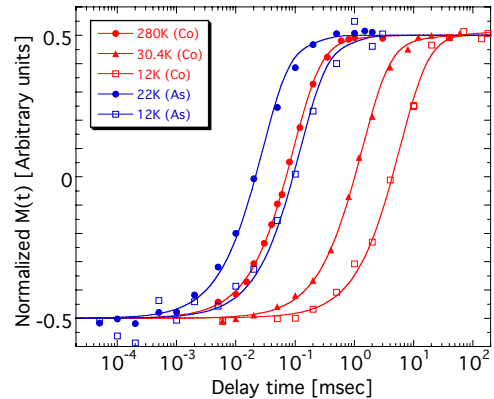}
\end{center}
\caption{(Color online) Typical nuclear spin recovery curves $M(t)$ after an inversion pulse for $^{59}$Co (red symbols) and $^{75}$As (blue symbols).  Solid curves are appropriate fits to determine $T_1$ with the solution of master equation for $I = \frac{7}{2}$ ($^{59}$Co) and $\frac{3}{2}$ ($^{75}$As).  Data points at $12$~K are somewhat scattered because the signal to noise ratio below $T_c$ is limited by the RF penetration depth and long $T_1$.}
\label{f4}
\end{figure}
Over all, both $^{59}(\frac{1}{T_1})$ and $^{75}(\frac{1}{T_1})$ in Fig.3 show qualitatively similar behavior as $^{19,75,139}(\frac{1}{T_1})$ measured at $^{19}$F, $^{75}$As, and $^{139}$La sites in the LaFeAsO$_{1-\delta}$F$_{\delta}$ superconductor ($\delta\sim 0.1$) \cite{ahilan1,nakai}.  We also note that the temperature dependence of $^{59}(\frac{1}{T_{1}T})$ is nearly identical to that of $^{57,59}K$.  The decrease of $^{59,75}(\frac{1}{T_{1}T})$ and $^{59,75}K$ below 280~K signals the suppression of the low energy spin excitations with decreasing temperature, i.e. {\it pseudo-gap behavior}.  We show the best fit to an activation type temperature dependence in Fig.2 and 3(b) with the same magnitude of a gap $\Delta_{PG}/k_{B} = 560$~K.  Our value is significantly larger than $\Delta_{PG}/k_{B} = 172$~K estimated for LaFeAsO$_{0.9}$F$_{0.1}$ \cite{nakai}.  

In the case of LaFeAsO$_{1-\delta}$F$_{\delta}$ or the oxygen deficient analogue,  LaFeAsO$_{1-2\delta}$, $(\frac{1}{T_{1}T})$ appears to level off in a very narrow temperature range below $\sim$50~K down to $T_c$.\cite{ahilan1,nakai,mukuda}  In the present case, both $^{59,75}K$ and $^{59}(\frac{1}{T_{1}T})$ completely level off below temperatures as high as $T^{*}\sim100$~K.\cite{orbital}  Interestingly, the in-plane electrical resistivity $\rho$ also shows a crossover to T-linear behavior, $\rho \sim T$ in the same temperature range.\cite{sefat3}  Even though our experimental results below $T^{*}$ satisfy {\it the Korringa relation} for canonical Fermi liquid, $(\frac{1}{T_{1}TK^{2}}) = constant$, the $\rho\propto T$ behavior is inconsistent with the canonical Fermi liquid picture.  Thus these findings suggest that a new electronic phase emerges below $T^{*}$ prior to the superconducting transition at $T_{c}=22$~K.

A closer look at Fig.3(b) reveals that $^{75}(\frac{1}{T_{1}T})$ increases with decreasing temperature toward $T_c$.  The origin of the different temperature dependence of $^{75}(\frac{1}{T_{1}T})$ from $^{75}K$ and $^{59}(\frac{1}{T_{1}T})$ is not clear, but possible mechanisms include the following: (a) Since $A<0$ and $B>0$, the hyperfine form factor at Co sites may satisfy  $|^{59}A({\bf q})|^{2} \sim 0$ for some intermediate wave vector values.  If spin fluctuations grow for these modes toward $T_c$, $^{59}(\frac{1}{T_{1}T})$ is insensitive to their enhancement while $^{75}(\frac{1}{T_{1}T})$ would grow toward $T_c$.  (b) Some of the Fe and Co 3d orbitals are orthogonal to 4s and 4p orbitals at $^{75}$As sites, and therefore spin transfer from these 3d orbitals is very limited.  This means that $^{75}(\frac{1}{T_{1}T})$ may be insensitive to spin fluctuations in certain Fe and Co 3d orbitals.  (c) The hyperfine form factor at $^{75}$As sites satisfy $|^{75}A({\bf q})|^{2}=0$ for wave vector ${\bf q}$ at the boundary of the first B.Z., and hence $^{75}$As sites are insensitive to spin fluctuations with large momentum transfers (i.e. commensurate antiferromagnetic modes).  In other words,  $^{75}(\frac{1}{T_{1}T})$ may be dominated by smaller wave vector modes of spin fluctuations in Co 3d orbitals that are not orthogonal to As orbitals.  Given that $^{75}K$ is temperature independent below $T^{*}$ toward $T_c$, however, we rule out the possibility of ferromagnetic enhancement of the ${\bf q} \sim {\bf 0}$ modes toward $T_c$.    

Finally, we turn our attention to the superconducting state below $T_c$.  In Fig.2, we present preliminary $^{75}K_{c,ab}$ data below $T_c$ measured with an external magnetic field $B = 7.7$~Tesla applied along the crystal c-axis or ab-plane.  {\it Our results show, for the first time, that both $^{75}K_{c}$ and $^{75}K_{ab}$ decrease below $T_c$}.  This finding is consistent with the singlet pairing of superconducting Cooper pairs, but in conflict with the p-wave triplet superconductivity model.  In the latter scenario, spin susceptibility  should remain constant below $T_c$ at least along one orientation, hence either $^{75}K_{c}$ or $^{75}K_{ab}$ should remain constant across $T_c$.   In passing, the decrease of $^{75}K_{ab}$ below $T_c$ was previously reported for (La,Pr)FeAsO$_{1-\delta}$F$_{\delta}$,\cite{grafe,zheng} but one needs to measure {\it both} $^{75}K_{c}$ {\it and} $^{75}K_{ab}$ to test the triplet superconductivity scenario.  In Fig.3, we observed neither the Hebel-Slichter coherence peak nor the exponential decrease of $^{59,75}(\frac{1}{T_1})$ expected for conventional BCS s-wave pairing.  Our results are  similar to earlier reports.\cite{nakai,grafe,zheng,mukuda}  Both $^{59,75}(\frac{1}{T_1})$ appear to level off near 4~K, presumably because of disorder in Fe$_{1.8}$Co$_{0.2}$As$_{2}$ layers and the influence of vortices.  At a qualitative level, the observed behavior of $^{59,75}(\frac{1}{T_{1}})$ below $T_c$ is analogous to that observed for high $T_c$ cuprates with d-wave pairing symmetry.\cite{imai}  

To summarize, we have reported the first NMR investigation of the electron-doped superconductor BaFe$_{1.8}$Co$_{0.2}$As$_{2}$ using a single crystal.  We showed that Co substitution does not induce local moments.  Instead, our results support the earlier proposal that the extra Co electrons become itinerant superconducting carriers in the Fe$_{1.8}$Co$_{0.2}$ sheets.\cite{sefat2,sefat3,german}  An inevitable conclusion is that the underlying physics of FeAs layers is quite different from that in high-$T_c$ cuprates, where the generally accepted starting point is the doped Mott insulator.  In this context, it is interesting to notice the similarity between the present case and the itinerant magnetic system Y(Co$_{2-x}$Al$_x$).\cite{yoshimura}  Alloying induces little modulation of local electronic properties in Y(Co$_{2-x}$Al$_x$), and $^{59}$Co and $^{23}$Al NMR show similar behavior.\cite{yoshimura}  On the other hand, our observation of the pseudo-gap behavior in this as well as LaFeAsO$_{0.89}$F$_{0.11}$ system\cite{ahilan1} suggests that it is a common property shared among the various iron-pnictide superconductors, as well as the high-$T_c$ cuprates. 

T.I. acknowledges financial support by NSERC, CFI and CIFAR.  Research sponsored by the Division of Materials Science and Engineering, Office of Basic Sciences, Oak Ridge National Laboratory is managed by UT-Battelle, LLC, for the U.S. Department of Energy under contract No. DE-AC-05-00OR22725.\\

\end{document}